\documentclass[letterpaper, 10 pt, conference]{ieeeconf}
\let\labelindent\relax% Comment this line out if you need a4paper

\IEEEoverridecommandlockouts                              % This command is only needed if 
\usepackage{booktabs}
\usepackage{amsmath}
\usepackage{balance}
\usepackage{enumitem}
\usepackage{hyperref}
\usepackage{amssymb}
\usepackage{colortbl}
\usepackage{stfloats}
\usepackage{xcolor} 
\usepackage{textcomp}
\usepackage{graphicx}
\usepackage{verbatim}
\usepackage{multirow}
\usepackage{algorithm}
\usepackage{multicol}
\usepackage{algpseudocode}
\usepackage{mathtools}

\usepackage{hyperref}
\usepackage{subcaption}

\graphicspath{{./pdf/}{./jpeg/}{./eps/}}

\algrenewcommand\algorithmicrequire{\textbf{Input:}}
\algrenewcommand\algorithmicensure{\textbf{Output:}}
\def\BibTeX{{\rm B\kern-.05em{\sc i\kern-.025em b}\kern-.08em
    T\kern-.1667em\lower.7ex\hbox{E}\kern-.125emX}}

\title{\LARGE \bf
Quantum Approaches for Dysphonia Assessment \\in Small Speech Datasets 
}

\author{Ha Tran$^{1}$, Bipasha Kashyap$^{1}$, \textit{Member, IEEE}, Pubudu N. Pathirana$^{1}$, \textit{Senior Member, IEEE}% <-this % stops a space
\thanks{$^{1}$All authors are with the Networked Sensing \& Biomedical Engineering (NSBE) Research lab, School of Engineering, Deakin University, Waurn Ponds, Victoria, AU.
        {\tt\small thi.n.tran@deakin.edu.au, aubipasha@deakin.edu.au, pubudu@deakin.edu.au}}%
}

\begin{document}

\maketitle
\thispagestyle{empty}
\pagestyle{empty}

%%%%%%%%%%%%%%%%%%%%%%%%%%%%%%%%%%%%%%%%%%%%%%%%%%%%%%%%%%%%%%%%%%%%
\begin{abstract}
Dysphonia, a prevalent medical condition, leads to voice loss, hoarseness, or speech interruptions. To assess it, researchers have been investigating various machine learning techniques alongside traditional medical assessments. Convolutional Neural Networks (CNNs) have gained popularity for their success in audio classification and speech recognition. However, the limited availability of speech data, poses a challenge for CNNs. This study evaluates the performance of CNNs against a novel hybrid quantum-classical approach, Quanvolutional Neural Networks (QNNs), which are well-suited for small datasets. The audio data was preprocessed into Mel spectrograms, comprising 243 training samples and 61 testing samples in total, and used in ten experiments. Four models were developed (two QNNs and two CNNs) with the second models incorporating additional layers to boost performance.
The results revealed that QNN models consistently outperformed CNN models in accuracy and stability across most experiments.

\indent \textit{Clinical relevance}— This work investigates the potential of quantum-based approaches for medical data classification and their promising role in enhancing dysphonia assessment.
\end{abstract}

\section{Introduction}
Speech is a fundamental aspect of human communication, facilitated by the coordinated function of various organs. However, these organs can be compromised by neurological disorders such as Parkinson’s, Ataxia, and Dysphonia, making communication challenging for many individuals. Dysphonia, a prevalent medical condition, affects approximately 10\% of the general population and 50\% of professional voice users~\cite{martins2016voice}. Those who use their voice extensively, such as teachers, or individuals in older age groups, are more prone to developing this condition. Symptoms of Dysphonia often include hoarseness, a weak or breathy voice, strained speech, or even voice loss, typically caused by malformations or dysfunctions of the vocal cords or the larynx (voice box).  

Dysphonia can be assessed by medical history, symptom assessment, and laboratory examination. Assessment can be done using the GRBAS (Grade, Roughness, Breathiness, Asthenia, and Strain) scale to assess voice quality ~\cite{hirano1981clinical}. Equipment such as a laryngoscope is used to examine the vocal cords and larynx. Imaging tests such as CT or MRI scans can be used to recognize structural or neurological abnormalities. However, these methods are time-consuming, discomfort for patients, and expensive. According to~\cite{cohen2012direct}, a patient needs to pay from 577 to 953 US dollars per year for diagnosing and managing the disorder.    

Machine learning has made significant advancements in analysing voice signals, which can be effectively used to assess dysphonia and other voice disorders. Some common features of voice signals such as Mel-Frequency Cepstral Coefficients (MFCCs), jitter, and shimmer are extracted as the input of models such as Random Forest, Decision Trees, Support Vector Machine (SVM), Gaussian Mixture Model Supervector Kernel-support Vector Machine (GMM-SVM)~\cite{verde2018voice, rehman2024voice, dankovivcova2018machine, wang2011discrimination}. Although these works achieved high accuracy in detecting Dysphonia, these ML algorithms require expertise and experience in feature selection and combination.  

Deep learning methods, such as Convolutional Neural Networks (CNNs), have been widely used in tasks like image recognition, speech classification, and natural language processing. These methods have also shown promise in assessing pathological speech by leveraging advanced feature extraction and classification capabilities. Specifically, in~\cite{islam2022voice}, CNN was used to classify the Dysphonia and normal voices and achieved an accuracy of 82.33\%. The study~\cite{wu2018convolutional}  achieved 88.5\%, 66.2\%, and 77\% classification accuracy on training, validation, and testing data using CNN. Despite these promising results, the application of deep neural networks to speech-related rare disorders is challenging due to the limited availability of medical data. Small datasets often result from the rarity of these conditions, and hospitals are often hesitant to share data due to privacy concerns. To address these challenges, researchers have explored techniques like oversampling, as demonstrated in ~\cite{lee2023efficient}, where CNNs with oversampling achieved an impressive accuracy of 98.9\%. Another approach ~\cite{peng2023voice} utilized a combination of pre-trained CNNs and SVM classifiers to enhance performance with limited data, using Mel spectrograms as input. The study~\cite{chen2023deep} divided audio files to augment the data and used CNN to classify Dysphonia and normal people based on MFCCs, and Mel spectrogram features. This framework obtained 92\% accuracy, 98\% recall, 89\% precision, and 87\% specificity. These advancements highlight the potential of deep learning in speech pathology but underscore the need for innovative solutions to overcome data scarcity and privacy barriers.

In recent years, quantum machine learning~\cite{schuld2021machine} has been an emerging field with various applications. Classical machine learning algorithms still suffer from computational bottlenecks such as model complexity, high dimensionality, and processing power. Some researchers show that quantum algorithms can perform well on small datasets. In~\cite{sagheer2019novel}, an autonomous perceptron model (APM) inspired by the computational power of the qubit outperformed some classical machine learning models in terms of accuracy and computational time with the limited number of training samples. In~\cite{acar2021covid}, quantum transfer learning was used to detect COVID-19 from CT images and achieved an accuracy of 90\% to 100\% using 20\% and 80\% of training and testing data. Relatively little study has been done in the field of quantum speech thus far.~\cite{qi2022classical} proposes a hybrid transfer learning approach combining classical CNN and Variational Quantum Circuit (VQC)-based Quantum Neural Network to enhance spoken command recognition on noisy intermediate-scale quantum (NISQ) devices, achieving improved performance on the Google Speech Commands dataset. ~\cite{hong2022qspeech} proposed a Quantum Neural Network with a low-qubit VQC and created a comprehensive application framework known as QSpeech. ~\cite{chen2024consensus} introduces a Consensus-based Distributed Quantum Kernel Learning (CDQKL) framework designed to enhance speech recognition by distributing computational tasks across quantum terminals connected via classical channels, thereby preserving data privacy and improving scalability.   In 2021,~\cite{yang2021decentralizing} proposed a distributed speech command recognition algorithm based on a Quantum Convolutional Neural Network (QCNN) or QNN, encoding Mel spectrograms of audio by using 2 × 2 quantum convolution layers, with subsequent networks being classical deep learning networks. The experiment results showed that the algorithm achieved an accuracy of 95.12\% on the Google Speech Commands dataset.

As per our literature review, there is currently no prior work on the benefits of Quantum approaches in detecting voice and speech disorders. This study is a novel attempt at hybrid quantum-classical algorithms, the Quanvolutional Neural Network (QNN)~\cite{henderson2020quanvolutional}, which combines the advantages of quantum and classical technology, to detect Dysphonia. The contributions of this paper are as follows.

1. Investigating the behaviour of the hybrid quantum model in detecting voice disorder.

2. Comparing experimental results and analysing the performances between CNNs and QNNs on small datasets.
\section{Methodology}
\label{method}
\subsection{Processing Speech}
\label{subsection: speech process}
\begin{figure}[htbp]
    \centering
    \includegraphics[width=0.5\textwidth]{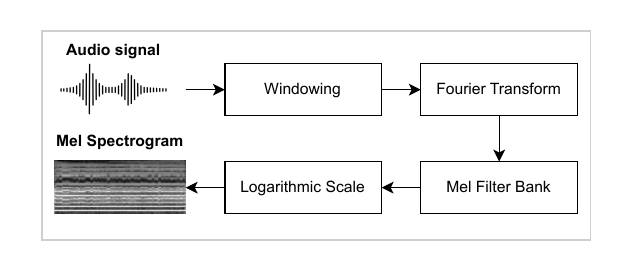}
    \caption{The speech processing.}
    \vspace{-10pt}
    \label{fig:speech process}
\end{figure}

A Mel spectrogram is a visual representation of frequencies over time of the audio signal. It is one of the common features used in fields such as speech recognition~\cite{park2019specaugment}, audio classification~\cite{xu2018large, meghanani2021exploration}. Figure~\ref{fig:speech process} shows the process of converting an audio signal into a Mel spectrogram. Specifically, the audio time-domain signal is divided into overlapping frames using a windowing function such as a Hamming or Hann window. Each frame is then transformed into a spectrogram representing frequencies over time using the Fourier Transform. The spectrogram is passed through a Mel filter bank and converted to a logarithmic scale in decibels (dB). The final output is a 2D Mel spectrogram image, where the time and frequency are on the horizontal and vertical axis. In this study, the Mel spectrogram is generated using the Librosa library with a window size of 2048, hop-length of 512 samples, 2048-point Fast Fourier Transform, 128 Mel bands, and Grayscale intensity to reduce the complexity. The final Mel spectrogram having the dimension of 40-by-100-by-1 (height-by-weight-by-channel) is used as the input of the QNNs below.

\subsection{Quanvolutional Neural Network}
QNN is the hybrid quantum-classical neural network. The quanvolutional layer replaces the convolutional layer in the classical CNN model to improve performance by extracting features using quantum circuit properties. These features are subsequently aggregated and passed to the following layers in the classical neural network, enabling further processing and classification.

\subsubsection{\textbf{Quanvolutional Layer}}
\label{subsub:Quanv layer}
\begin{figure}[htbp]
    \centering
    \includegraphics[width=0.5\textwidth]{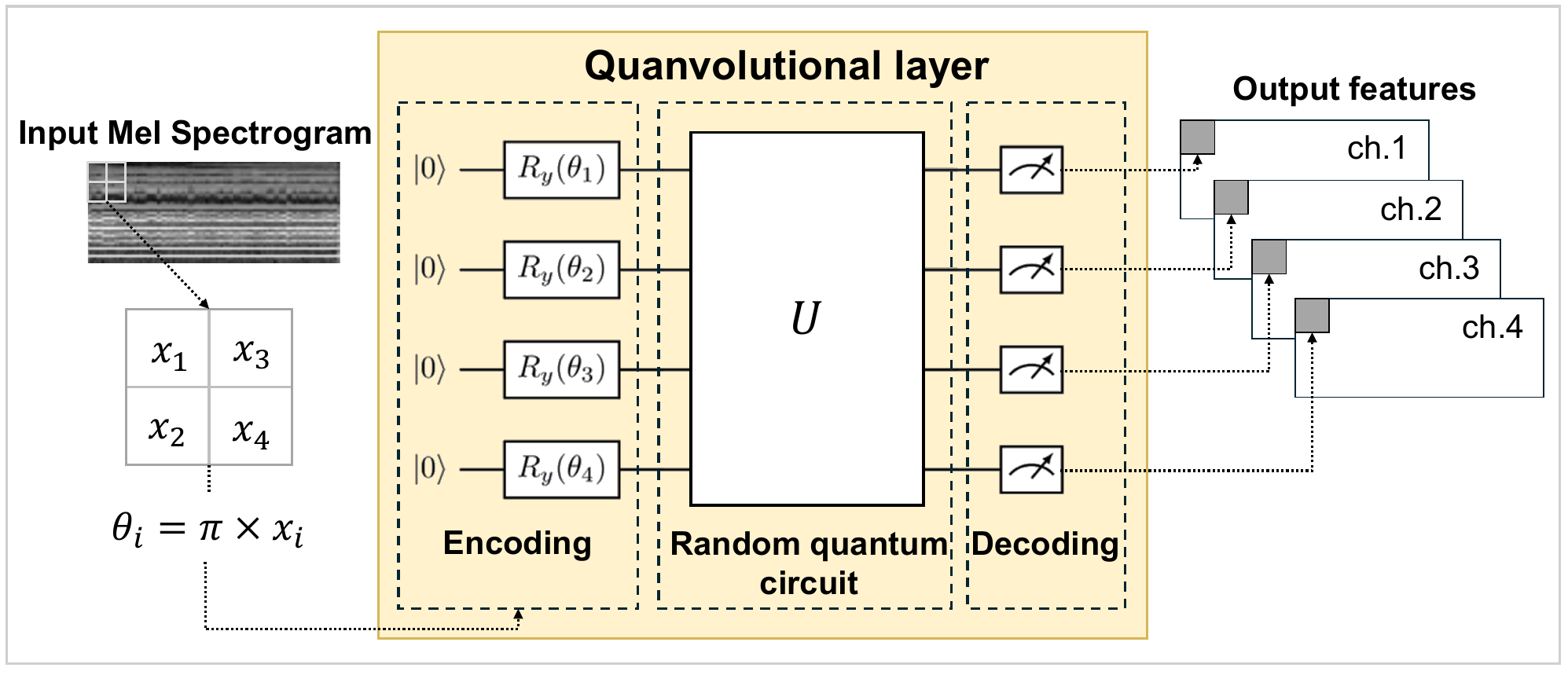}
    \caption{The quanvolutional layer.}
    \vspace{-10pt}
    \label{fig:Quanvolutional layer}
\end{figure}

A quanvolutional layer, which contains many quanvolutional filters, transforms patches of the input tensor using quantum circuits instead of performing element-wise matrix multiplication like the classical convolutional layer. These circuits, which can be structured or random, process the data by utilizing quantum properties such as superposition and entanglement. In this study, we use a 2-by-2 quanvolutional filter, corresponding to four qubits and a random quantum circuit. Figure~\ref{fig:Quanvolutional layer} illustrates the three main parts of the quanvolutional layer, including Encoding, Random quantum circuit, and Decoding.
\paragraph{Encoding}
There are several encoding methods including basis, amplitude, and angle encoding. In this research, angle encoding is selected using parametrized rotations ($R_y(\theta)$) four initialized qubits in the ground state. A 2-by-2 square of the Mel spectrogram image is used as the input of the encoding part. Four rotational parameters $\theta_i$ are calculated from the corresponding four intensity values of the 2-by-2 square of the input image, and scaled by a factor of $\pi$. Therefore, the classical input data is encoded into quantum data. 
\begin{figure*}[!htbp]
    \centering    
    \resizebox{0.9\textwidth}{!}{
    \includegraphics[width=1\textwidth]{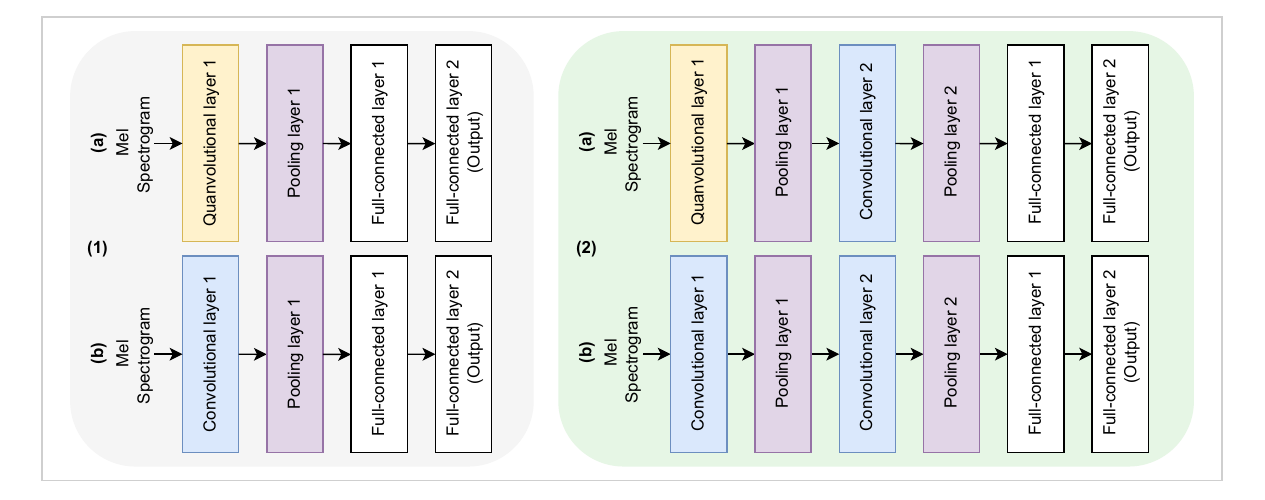}}
    \caption{The proposed framework: (1a) Quanvolutional Neural Network 1 (QNN1)  and (1b) Convolutional Neural Network 1 (CNN1), (2a) Quanvolutional Neural Network 2 (QNN2) and (2b) Convolutional Neural Network 2 (CNN2). The (1) gray, and (2) green  boxes represent the models used in the first and second experimental scenarios, respectively. The only difference between CNNs and QNNs is the replacement of the convolutional layer by the quanvolutional layer.}
    \vspace{-15pt}
    \label{framework}
\end{figure*}
\paragraph{Random quantum circuit}
A quantum circuit, denoted as $U$, extracts features from encoding data. The random quantum circuit uses random quantum gates and parameters. Specifically, some 1-qubit gates ($R_x(\theta),R_y(\theta),R_z(\theta),T,H$) and 2-qubit gates ($CNOT$, $SWAP$ or $CZ$) are chosen randomly and parameterised with random angles ($\theta$). By generating entanglement between qubits, multi-qubit gates enable the circuit to take advantage of quantum correlations that are difficult for classical systems to reproduce. 

\paragraph{Decoding}
Decoding is the process of converting the output of the quantum circuit into classical data, which is the input of the following classical layers. The quantum states are measured to obtain classical values. Some common measurements include expectation values and probability distributions. In this research, we employ Pauli-Z gates to directly use the raw expectation values.

After measurement, four expectation values are mapped to four corresponding channels (e.g. ch.1, ch.2, ch.3, ch.4 in Figure~\ref{fig:Quanvolutional layer}). Thus a 2-by-2 square input image convolved into a single output pixel. Repeating this procedure over the whole input image yields a multi-channel output image. Classical layers or additional quantum layers can follow this quanvolutional layer. 

\subsubsection{\textbf{Quanvolutional network}}
The quanvolutional layer can be seamlessly embedded into various architectures, similar to classical convolutional layers. Users are required to define the number of filters in each layer, the total number of layers, and their arrangement within the model. For instance, as demonstrated in~\cite{vu2024exploring}, a QNN architecture may include one quanvolutional layer followed by two fully connected layers.
Another example~\cite{henderson2020quanvolutional}, the quanvolutional layer is the first layer, which is followed by a pooling layer, a convolutional layer, a second pooling layer, and two fully-connected layers. In this study, we investigate the performance of quanvolutional layer when combined with only fully-connected layers and convolutional layers, as described in section~\ref{tested_models}.

\section{Experiments}

\subsection{Dataset}
The speech dataset is the Perceptual Voice Qualities Database (PVQD)~\cite{walden_pvqd_2020}. This dataset consists of 296 raw audio files containing the sustained /a/ and /i/vowels and the sentences from Consensus Auditory-Perceptual Evaluation of Voice. We used the vowel /a/ in the dataset to assess the viability of the proposed framework. The audio segments corresponding to vowel /a/ are extracted from the raw audio files. The resulting dataset includes 216 and 88 audio files of Dysphonia patients and Healthy people, respectively. This speech dataset would be converted to the Mel Spectrogram image dataset, as described in section~\ref{subsection: speech process}, and then split into 243 training and 61 testing images. In order to prove our hypothesis about the outstanding QNN with a small dataset, the models would be trained with some experiments and tested in all 61 testing samples. Specifically, there are 10 experiments with an increasing number of training patterns: 60, 80, 100, 120, 140, 160, 180, 200, 220, and 240. These samples of each experiment are randomly selected with the same ratio between Dysphonia and normal people (70/30) as the raw dataset.
\subsection{Models}
\label{tested_models}
The performance of the quanvolutional layer in QNNs is compared against that of classical CNNs with standard convolutional layers. In this study, we evaluate two scenarios, as illustrated in Figure~\ref{framework}. First, we test the simplest form of a QNN. Next, we incorporate convolutional and pooling layers, following the approach outlined in~\cite{henderson2020quanvolutional}.

\subsubsection{\textbf{Scenario 1}}
The simplest quanvolutional neural network model, QNN1, contains: Quanvolutional layer 1 (QUANV1), as described in section~\ref{subsub:Quanv layer}, Pooling layer 1 (POOL1), Full-connected layer 1 (FC1), Full-connected layer 2 (FC2). The corresponding classical convolutional neural network 1 model, CNN1, replaces the first QUANV1 by the Convolutional layer 1 (CONV1), and then remains the following structure of QNN1. The relevant CONV1 layer used ReLU activation function, 2-by-2 kernel size, valid padding and 4 filters. The POOL1 layer with the shape 2-by-2 reduced the dimension by 2. The data was flattened and put into the dense block at the end of the POOL1 operation. This block consisted of 64 fully connected layer (FC1) with tanh activation function and 1 fully connected layer (FC2) with softmax activation function, with a dropout probability of 0.5. 
\subsubsection{\textbf{Scenario 2}}
In the second scenario, two layers including the Convolutional layer 2 (CONV2), and the Pooling layer 2 (POOL2) were added between the POOL1 and FC1 of the first case model, as shown in Figure~\ref{framework}. The CONV2 layer has 16 filters, 2-by-2 kernel size, ReLU function, and same padding. The POOL2 layer also halves the dimension of the image. The second quanvolutional and classical neural network models, QNN2 and CNN2, have the same first quanvolutional (QUANV1) and classical convolutional (CONV1) layers as the first scenario. The following layers are the same as the first scenario. 

By comparing the QNN model to the CNN model, we can address whether quantum features outperform the classical layers or not, and investigate the QNN performance when combined with the classical convolutional layers. PennyLane~\cite{bergholm2018pennylane}, which is an open-source software library for quantum computing and quantum machine learning, is used to simulate the quanvolutional layer in the classical computer. The models are constructed using version 2.15.0 of Tensorflow and 0.38.0 of Pennylane. The hardware used to train the models is a computer with a 6 cores AMD Ryzen 5 9600X CPU and NVIDIA RTX 4060 Ti GPU with 8GB DDR6 VRAM. Early stopping ~\cite{prechelt2002early} and k-fold Cross-Validation are used for training to ensure the models are generalized well. 
Each model is trained for 3000 iterations and 10 training steps, with Early Stopping applied to halt training if the testing loss does not decrease after 15 consecutive epochs. Moreover, the mean and standard deviation of testing accuracy are calculated after training 10 folds, and plotted as the line and shaded regions in Figure~\ref{fig:Acc_training}, respectively.

\section{Results and Discussion}
\label{sec:result, disscuss}
\begin{figure}[ht!]
    \centering
    % Top-left plot
    \begin{subfigure}[b]{0.5\textwidth}
        \centering
        \includegraphics[width=\textwidth]{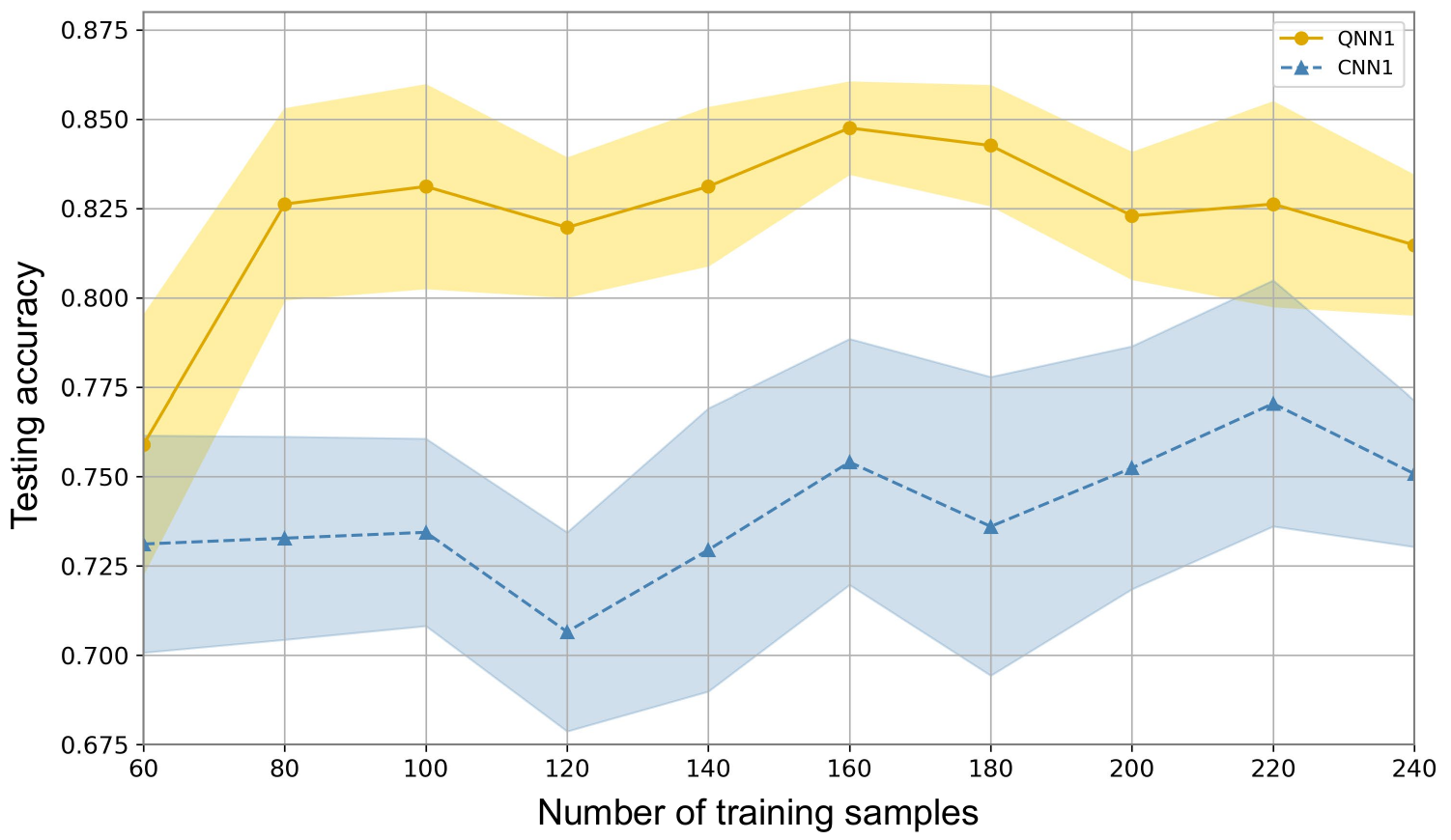}
        \caption{Comparison of testing accuracy between QNN1 and CNN1.}
        \label{fig:acc_training_1}
    \end{subfigure}
    % Top-right plot
    \hspace{0.3cm} % Add small horizontal space
    \begin{subfigure}[b]{0.5\textwidth}
        \centering
        \includegraphics[width=\textwidth]{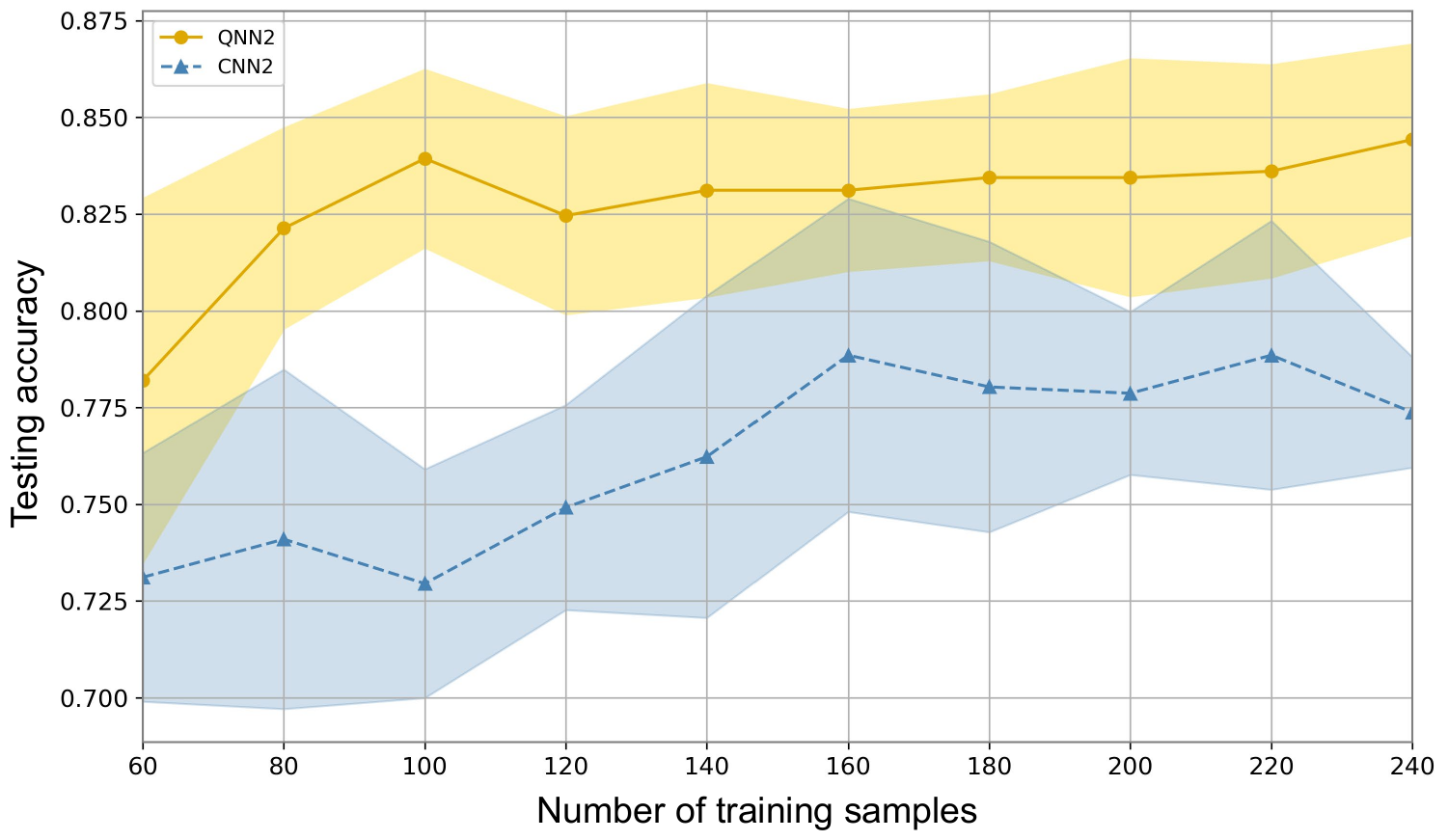}
        \caption{Comparison of testing accuracy between QNN2 and CNN2.}
        \label{fig:acc_training_2}
    \end{subfigure}
    \caption{Comparison of testing accuracy of (a) 1st and (b) 2nd scenario.}
    \label{fig:Acc_training}
    \vspace{-10pt}
\end{figure}

\begin{figure}[htbp]
    \centering
    \includegraphics[width=0.48\textwidth]{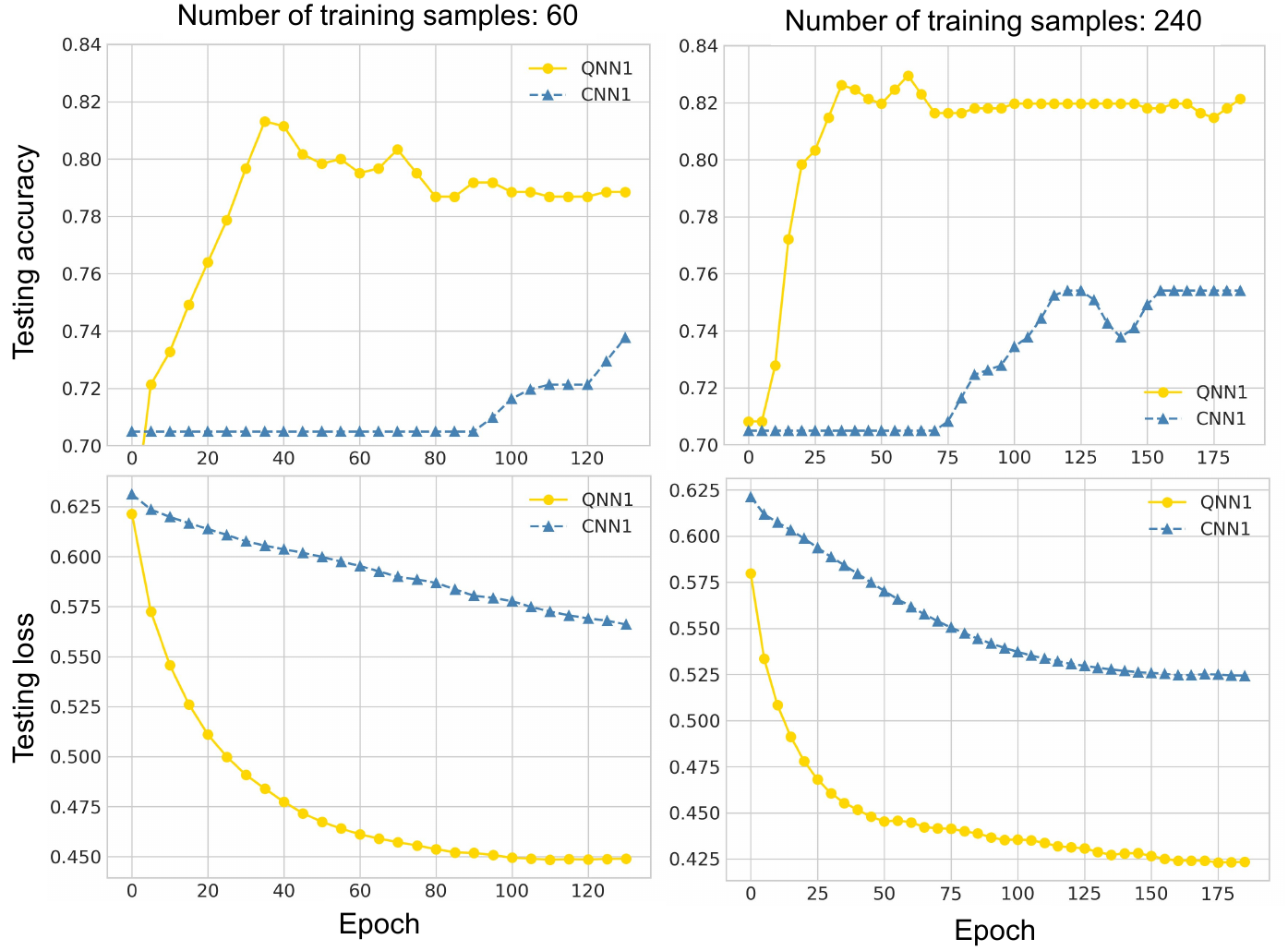}
    \caption{Comparision of testing accuracy (\textit{above}) and testing loss (\textit{below}) of QNN1 and CNN1 at 60 and 240 training samples.}
    \label{fig:Acc_loss_1}
    \vspace{-10pt}
\end{figure}

\begin{figure}[htbp]
    \centering
    \includegraphics[width=0.48\textwidth]{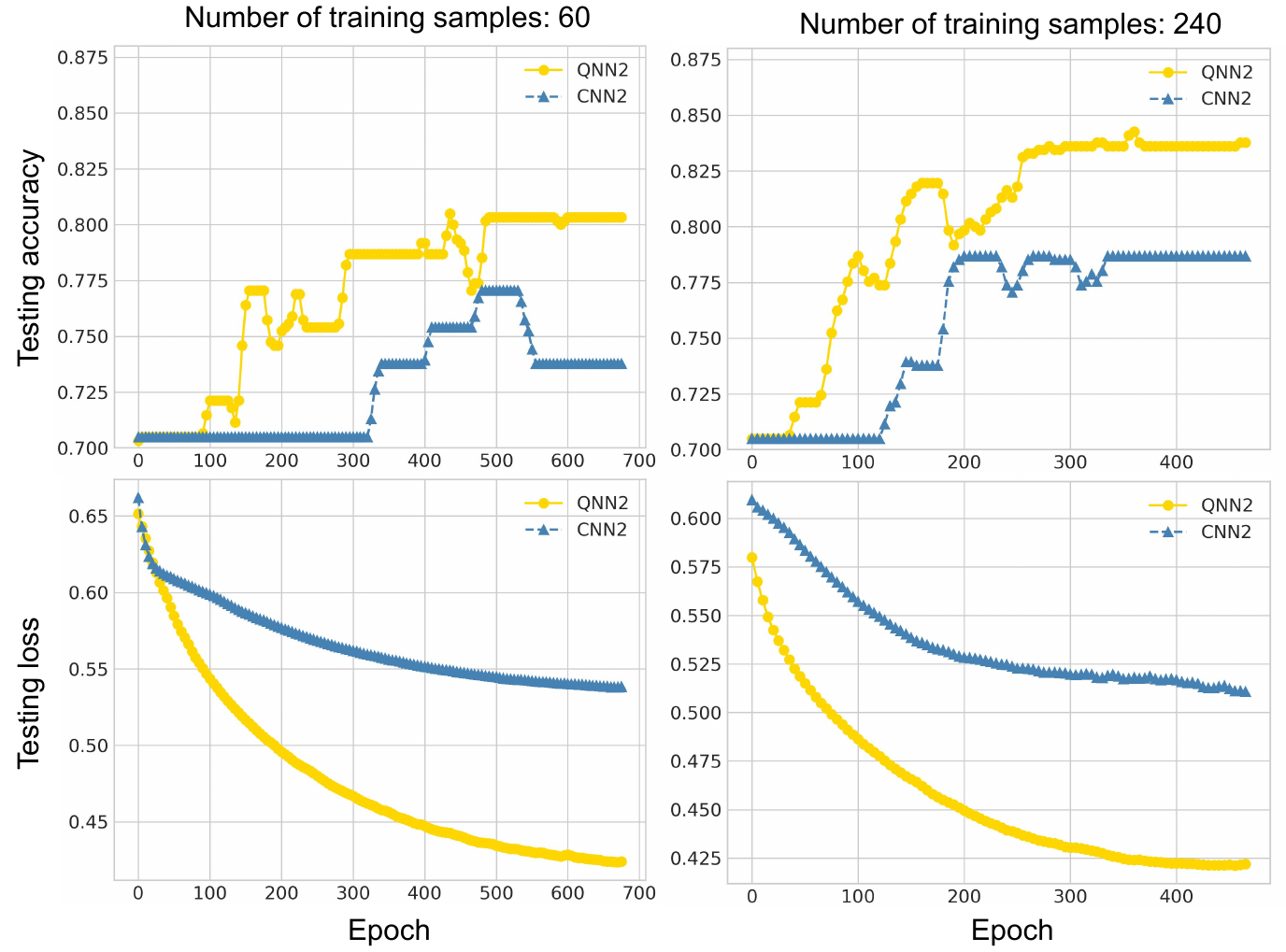}
    \caption{Comparision of testing accuracy (\textit{above}) and testing loss (\textit{below}) of QNN2 and CNN2 at 60 and 240 training samples.}
    \label{fig:Acc_loss_2}
    \vspace{-10pt}
\end{figure}

To evaluate how these models perform with a small training dataset, Figure~\ref{fig:Acc_training} compares QNN and CNN models across different numbers of training samples in two scenarios, as outlined in section~\ref{tested_models}. The QNN models (QNN1 and QNN2) consistently demonstrate higher accuracy and smaller standard deviations compared to the CNN models (CNN1 and CNN2), particularly with limited training samples. Specifically, in Figure~\ref{fig:acc_training_1}, for small training sample sizes (60-160), QNN1 achieves a mean classification accuracy of 76\%-85\%, compared to 73\%-75\% for CNN1. QNN1 maintains a more stable performance with consistently accurate predictions across all sample sizes, whereas CNN1 exhibits greater variability and larger standard deviations, indicating higher fluctuations in performance. In Figure~\ref{fig:acc_training_2}, QNN2 consistently outperforms CNN2 across all training sample sizes, maintaining higher accuracy. The standard deviations further show that QNN2 has a smaller variance, while CNN2 experiences more significant fluctuations in accuracy. These results suggest that QNNs can better leverage quantum properties for learning, making them a promising alternative to classical approaches, particularly for tasks involving limited or noisy data. Figures~\ref{fig:Acc_loss_1} and ~\ref{fig:Acc_loss_2} further evaluate the performance of QNN and CNN models on two experiments corresponding to 60 and 240 training samples, respectively. These figures provide a detailed comparison of accuracy and loss across the number of epochs to assess the models' performance for the two different dataset sizes. QNN1 and QNN2 attain higher accuracy more quickly than CNN1 and CNN2. The loss comparison reveals that QNNs stabilize and reduce loss more efficiently, underscoring superior performance.

Compared to QNN1, QNN2 is more accurate when two extra layers are added. In particular, QNN1 and QNN2 have approximate classification accuracy of 76\% to 86\% and 78\% to 87\%, respectively. From 140 to 240 training samples, the accuracy of QNN2 remains relatively constant, compared to that of QNN1. These results demonstrate the viability of combining the convolutional and quanvolutional layers to enhance classification.  

\section{Conclusion}
In this paper, we used the Quanvolutional Neural Network model to detect Dysphonia using Mel spectrograms. By comparing its performance with a classical convolutional neural network (CNN) across an increasing number of training samples, we were able to evaluate the effectiveness of QNNs. Our experimental results demonstrated that even with a limited dataset, QNNs achieved higher accuracy compared to CNNs. This study underscores the potential of integrating classical convolutional and quanvolutional layers to enhance classification accuracy.

There are, however, areas for improvement that will guide our future research. For instance, we can diversify the training data using techniques such as adding noise, time-stretching, shifting, and pitch alteration. Furthermore, experimenting with different encoding and decoding approaches may further enhance performance. Additionally, integrating quantum layers with other classical models will be explored to optimize the QNN framework.

\section*{Acknowledgement}
This research is supported by Network Sensing \& Biomedical Engineering (NSBE) Research Lab, School of Engineering, Deakin University. 
\bibliographystyle{IEEEtran}
\bibliography{Paper}

\end{document}